# On Universal Physical Reality in the Light of Quantum Consciousness

[1]Pabitra Pal Choudhury, [2]Swapan Kumar Dutta, [3]Sk. Sarif Hassan and [4]Sudhakar Sahoo

[1,2,3]Applied Statistics Unit, Indian Statistical Institute, Kolkata, 700108, INDIA

Email: pabitrapalchoudhury@gmail.com, sarimif@gmail.com

[4]Department of Computer Sc., Silicon Institute of Technology, Patia, Bhubaneswar-751024

Email: sudhakar.sahoo@gmail.com

**Abstract:** In this paper, we have first given an intuitive definition of 'Consciousness' as realized by us. Next, from this intuitive definition we derived the physical definition of quantum consciousness (Quantum Consciousness Parameter or QCP). This QCP is the elementary level of consciousness in quantum particles, which are the most elementary particles in nature. Thus QCP can explain both the perceptible and non-perceptible nature and some existing postulates of physics. We conceptualize that the level of human consciousness is most complex having highest fractal dimension of 4.85 in the electroencephalographs experiment done by other research groups. On the other hand, other species are having lesser consciousness level, which can be reflected by lesser fractal dimensions. We have also explored the bioinformatics of consciousness from genome viewpoints where we tried to draw an analogy of neurons with electrons and photons. Lastly, we refine the quantum mechanics in terms of QCP; we all know that in Einstein's special theory of relativity, Einstein has used the postulate "Consistency of the velocity of light irrespective of all frames of reference (inertial or non-inertial frames)". In our theoretical revelation QCP can be directly applied to get a confirmatory proof of this postulate. Thus the postulate can be framed as a law.

## 1. Introduction

*"Ohng Purnomadah Purnomidang Purnat Purnomudachyate
Purnosya Purnomadaya Purnomebaba Sisyate".*

The Sanskrit quotation says that whatever can / cannot be sensed by our sensitive organs and sophisticated instruments is building up the super-consciousness, which means as the following…

*Matter is transformed in finer form* $\rightarrow$ *Energy, is transformed in finer form* $\rightarrow$ *consciousness (reality up to quantum particle level) is transformed in finer form* $\rightarrow$ *Super consciousness (Bramha extending up to infinity).*

The spiritual background or true civilization of our country is at least five thousand years old. Our humble feeling is that there is no conflict between spirituality and science. Why we are saying so because our World Wide Web (WWW) is hardly 20 years old, now we know through the internet that we can exchange a piece of information across the world within a fraction of a second. So far, consciousness was considered to be a spiritual paradigm only because Physicist of the past were not sufficiently inclined on the subject, but consciousness is well within the context of science because it is a very fundamental reality within each of us.

It is observed that thinkers from all walks of life have contributed in some form or other in the understanding of consciousness. According to the meaning given in concise oxford dictionary, the term is defined as awareness, knowledge etc. Indian philosophers coined the term **"Chetana"** (meaning consciousness) from the Sanskrit origin **'Chit'** meaning knowledge.

Let us utter with the spiritual saying **"Sat-Chit-Anand"**.

**"Sat"** means absolute existence, which indicates about the rest mass (hypothetical), which cannot be subjected to any change with respect to Space-time and causality. **"Chit"** means absolute knowledge meaning the undistorted or pure knowledge and **"Anand"** means bliss meaning the joy everywhere and every-time. All events of our physical existence of nature are subjected to creation, expansion and termination. This is one part of consciousness. Again other part which is non-manifested or which cannot be sensed by so called "sensory organs or instruments" forms the basis of consciousness paradigm consisting of Mind, Intellect, Ego which are the things so far scientifically unexplored. We claim in this paper that both the perceptible and non-perceptible nature can be explained by quantum consciousness parameter (QCP) introduced by us. The thing is that although quantum mechanics with the present state of the art cannot distinguish between living and nonliving, QCP will form very much a complete story to switch from living to non-living and vice-versa explaining the fact of transition from lower entropy to higher entropy and vice versa. Under Physicist's terminology, QCP is derived as the impulse of energy which is again can be thought as momentum times energy or (mv) ($mc^2$) or $m^2vc^2$, obviously a vector quantity. This can be thought as the cause of any phenomenon and whenever this quantity is divided by Plank's constant, this quantity is the same as the Force acting on that event.

Consciousness is being studied in many different angles. We first dwell on the dimension of consciousness in section 2.1 introducing preliminary notions from Physics. Here we try to conceptualize the consciousness having more than three dimension and propose the fact that more the species are conscious the more their consciousness dimension. In section 3, an effort has been given to explain a small constituent of consciousness with the help of biological neurons. In this regard one project will be started soon at our Institute. We discuss in section 4, the chronological developments in Physics up to the present state of the art leading hopefully to Consciousness and in section 5 we redefine the quantum mechanics with the introduction of QCP. We all know that in Einstein's special theory of relativity, Einstein has used the postulate "Consistency of the velocity of light irrespective of all frames of reference (inertial or non-inertial frames)". In our theoretical revelation we beg to differ with that. Section 6 is the conclusion, which also highlights the future research directions on Consciousness.

## 2. Dimension of Consciousness

### 2.1. Concept of non-perceptibility of consciousness

In the previous section we have just introduced that momentum time energy of a quantum particle of mass m etc. constitutes the consciousness, the synergy of its full blossom will be seen in section 5. Thus from (mv), ($mc^2$) or $m^2vc^2$, we observe the dimension as $M^2L^3T^{-3}$. When this amount is divided by Plank's constant h, which is in Joules.sec equivalent to $ML^2T^{-1}$, we get $MLT^2$, which can be easily interpreted as the force of consciousness.

In the same paradigm, the velocity of the particle v = L/T = $\dfrac{h/mv}{h/mc^2}$ = $c^2/v$ pointing to the fact that the quantum particle can attain more than the speed of light and hence the consciousness becomes non-perceptible to the external world although it is a very basic fundamental reality.

### 2.2. Fractal dimension of consciousness

In 1904, one advancement in Science by Koch took place. Adding triangles to the sides of triangles ad infinitum produced enough mathematical curiosity. The resulting length of the triangle becomes infinitely long, yet remains contained in a finite space. In 1970, the word **"fractal"** was coined to describe fractions of dimensions.

Like the Koch snowflakes, fractals in nature maintain their irregular but distinctive shapes over different scales of magnification, which is known as nesting. Famous mathematician B. Mandelbrot was highly fascinated by the shape of the cauliflower- smaller and smaller pieces demonstrated self-similarity.
The fractal dimension of the outline of a typical cloud is 1.35. The fractal dimension of the coastline is 1.26. The fractal dimension of a piece of paper crumpled up into a ball is about 2.5.

Naturally, the fractal dimensions between 1 and 2 measures how wrinkly a line is. The crumpled paper ball fails to completely fill its allotted space, so it scores a dimension less than 3. Foamy structures may also have fractal shapes. The universe itself has a fractal shape due to its foam like structure caused by enormous globular voids between clusters of galaxies. In Biology it is seen in the sequential branching of ferns, trees, blood vessels, bile ducts and kidney (urinary collecting systems) into smaller and smaller versions of the original. Our vascular system, when stripped of all other cells, would almost fill the space that our bodies occupy. It is estimated to have a fractal dimension of slightly less than 3.

Computers may generate beautiful fractals when certain numbers are inserted into selected simple equations. When the results are fed back into the original equation, a feedback loop is set up, that is the mathematics feeds upon (recursion) its own results. We know already that feedback loops often have the effect of signal amplification as illustrated by a microphone being placed too close to its own loudspeaker. In 1940s, Donald O. Hebb investigated feedback loops in the brain (Mc Gone J. Going Inside. London: Faber and Faber, 2000: 43). He realized that successive firing in the same neural loop led to the reinforcement of nerve cell connections so that subsequent activation occurred more easily. This is as if the nerve networks have memory.

Surely, fractals and feedback loops are part of a branch of mathematics called nonlinear dynamics (ND). In the past, ND has been used to describe complex processes such as the weather, fluid turbulence, and many aspects of biology (mainly working of the mammalian brain), but still the brain function and consciousness remains elusive (see http:/mcb.berkley.edu/faculty/NEU/treemanw.html).

Next we observe the phase space of a simple pendulum as shown in fig-1.

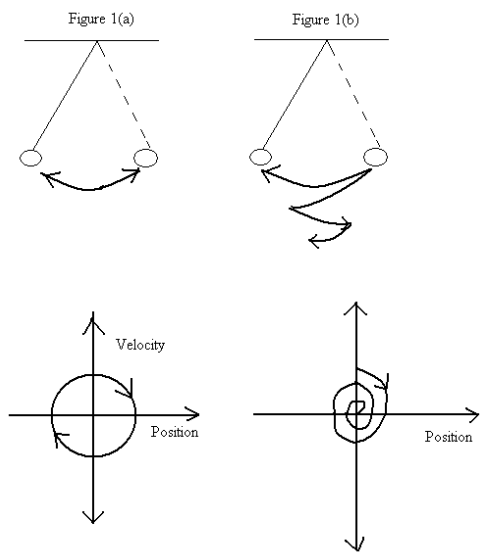

Figure 1: Phase space of a simple pendulum

The velocity and position of a pendulum is plotted to produce a circle in phase space. At any instant in time, the velocity and position collapse to a single point somewhere on the circle. The continuously scribed circle is a periodic attractor. The pendulum on the right has lost power, that's why the phase space spirals towards a point attractor. In contrast, for many biological systems, the attractor is the isoelectric state or cell death. In some nonlinear systems, unlike the case of a simple pendulum, the attractor is a fractal. One of the best known and first discovered fractal attractors associated with nonlinear dynamics is the pair of butterfly wings generated by a computer when Edward Lorent was studying equations relating to weather turbulence.

In biological systems, experimental data are often with noise and other artifacts. Thus it is often hard to discern the underlying dynamic processes.

It is to be noted that, the higher dimension involved in string theory are physical and thought to be highly compact. Phase space is a mathematical construct, and motion on an attractor is abstract. In fractal attractors the orbits in phase space may be stretched and folded like baker's dough when it is needed. In either case human imagination is not equipped to contemplate these higher dimensions directly. In other words, if data from perceptual space is retrieved, analyzed and shown to exhibit an underlying mathematical order that co-relates sensibly to its source, it is probably real even if it cannot be touched.
To look for signs of nonlinear dynamics, the data from electroencephalographs of different species including human are shown. Really an increase in mathematical sophistication across species is co-related with perceived notions regarding evolutionary rankings in the central nervous systems of the animals.

Table 1: Shows the data from electroencephalographs of different species including human

| SPECIES | FRACTAL DIMENSION |
|---------|-------------------|
| Human | 4.85 |
| Dog | 4.63 |
| Butterfly | 3.71 |
| Catfish | 2.50 |
| Crayfish | 1.65 |
| Earthworm | 0 |

Thus it is the firm conviction of the authors that appearance of consciousness is a revolutionary development during the evolutionary process. In humans a 4.8 dimension attractor may combine more than 4 variables in phase space. It is quite possible that the synthesis necessary for an explanation of consciousness occurs in the phase space associated with the nonlinear dynamics of the brain. It may be undeniably argued that objects in physical space enter our perceptual space via phase space. A hyperspace is a phase space with more than three dimensions (Churchland PS, Neurophilosophy: Towards a Unified Science of the Mind-Brain, Cambridge, MIT Press, 1986, 420-423.)

In [6, 7] we have given different possible applications of a newly developed model by us, which we named as Carry Value Transformation (CVT). Giving one fractal application in [6] we have highlighted that like other mathematical transformations CVT can also be used as an efficient tool to simulate the behavior of the nature. Mathematically we have also proved that just like Cellular Automata and Random

Boolean Network, CVT and Modified CVT (MCVT) both are discrete deterministic dynamical system and the cells (or values) in the table can be generated in parallel. In [7] the construction of CV Tables in different dimensions are highlighted. Also like CVT and MCVT their inverse transformation named as Extreme Value Transformation (EVT) is used to form different number theoretic fractals.

In a more general way, like CV table, MCV table and EV table any "table" with respect to a function or a relation (values of the table are generated with the help of this relation) also can generate both periodic as well as chaotic patterns. We can conjecture that one link can be established between these complicated patterns generated by this relational table with the theory of machine consciousness. More specifically identifying a suitable relation, choosing both rows and column numbers (integer, real etc.) in the table a chaotic and complex pattern can be generated which can mimic the complicated human brain. To map the exact brain structure a set of functions/relations may be non-uniformly (or hybrid) applied to the row and column numbers in the table.

Then, analysis of this table values may help us to understand the complicated structure of the brain and the function (sometimes operator or Rule) may be treated as the main logic, which is responsible for functioning of our mind. Again on changing different relation(s) in different time steps at time $t=0$, $t=1$, etc. dynamically table values can be changed and thus may be treated as another tool for evolutionary computing. Thus different thoughts or the changes of mind in different time instants can be suitably modeled using this relational table considering as a tool. In addition to this, other mathematical tools like Cellular Automata, Random Boolean Networks, L-Systems etc. are other possible options because these are also used to generate self-similar and chaotic fractals.

## 3. Research Effort on One Experiment on Consciousness

We know that in any digital computer, someone has to wire a computer and program it. But our brain's wiring consists fundamentally of laying down pathways of information flow and networks of information processing. Brain wiring is carried out by processes that are derived from the genome, but the information for this wiring is not entirely contained within the genome itself. Rather, the brain develops by a progressive sequence of steps that each involves interplay between genetic programs and environmental influences.

This view has antecedents (M. Sur, The Brain and Mind, 24$^{th}$ March 2009). The physicist Erwin Schrödinger, in his book "what is life" published before the discovery of the structure of DNA, pointed out the necessity of a crystalline genetic material in living organisms that would form the template for reproduction. The biologist Sydney Brenner has pointed out that the defining feature of a living organism is its ability to clone or recreate itself from its genetic code. We add that the brain wires itself from coded instructions that are open to and interpret extrinsic information, and this is in fact central to understanding human cognition and biological intelligence.

The earliest events in brain developments consist of neurons being and migrating to the right place. Subsequently, cells in one region of the brain extend axons or wires that link that region with another. Thus the eyes project to the visual cortex. These projections are highly specific- there are hardly any mistakes tolerated during such pathway development—and evolution has set up a large number of genes that orchestrate this specificity. In other words, Nature is highly deterministic in all cases of specificities. For example, that we are speaking now on a specific subject implies a specific set of networks etc. is definitely involved for such activations. This theme of specificity will be particularly applicable in section 5 where we elaborate that when the third party (observer) observes some event it might be associated with certain uncertainties, which afterwards could be removed with the introduction of our deterministic theory leading to the case of specificity.

The human genome project has transformed our ways of mapping genomic sequences and identifying genes, and has turned biology into an information science. Of the approximately 20,000 genes that comprise the human genome, about 80% are expressed in the brain. Different genes are expressed in different parts of the developing nervous system, and at different times. Sometimes the same gene is expressed in different brain regions, and sometimes a gene stays on for variable durations in different regions, but rarely is a gene turned off and then on again in the same region. Each gene thus has a unique function in space and time in the developing brain, but one that is influenced by context.

One of the most remarkable discoveries of the recent years is that genes can be regulated in a large number of ways. They can be turned off and on by critical control elements, including epigenetic mechanisms in the genome and small pieces of RNA. And not only what protein a gene makes but also how much it makes is influenced by a host of factors inside and outside the genome.

For communicating between two living elements some language is essential. This language may be in the very primitive terms look like sensing something. Humans have five sense organs through which transmission of information takes place. Let us consider the smell-sensing organ: Olfactory Receptors (ORs) which are found in the vertebrate genome as clusters of ~2-5 receptors per loci and have expanded via gene duplication events over evolution throughout the genome. It is still not known what a basic unit of these receptor cluster is, and how these building blocks propagate in the genome. Our research efforts will be focused on the clustered DNA and protein sequences with the objective of understanding the functional role of these clusters with respect to olfactory compounds. Also it is the conviction of the authors that QCP will act the pivotal role in framing the corresponding theories.

Basically, we are in search of Bio-informatics describing human genome and its role in consciousness.

Our prime objective is to characterize one small constituent of consciousness where biological neurons seem to take place of electrons and photons as found in electronic computers and optical computers respectively. Remember that traveling electrons are the constituents in an atom. But photons are different which are called bosons, whereas electrons are fermions. But Scientists are already aware of the dual existence of this kind of particles. In this context it may be highlighted that there exits various states in Physics, viz. solid, liquid, gaseous, plasma and the fifth state is definitely non-perceptible signifying the quantum particles' velocity greater than the velocity of light.

It is well known that classical mechanics, based on Newton's laws of motion, can predict the motion of planets around the sun accurately. In a similar way, it can also predict the motion of satellites around the earth. We know that a satellite can be launched with the help of a rocket at any desired height above the earth. Thus a satellite moving around the earth can occupy any orbit. In other words, all the orbits are permissible. A satellite in each of these orbits has an energy, which can be calculated by using the laws of classical mechanics. This means that there is no restriction on the energy of a satellite orbiting around the earth or a planet orbiting the sun. However, the same is not true for an electron inside the *atom*. Although the electrons are similar to planets or satellites in a much smaller system like the atom, things are quite different there.

In any atom, an electron can only occupy a set of ***permissible*** orbits around the atomic nucleus having ***discrete*** energy levels. Laws of classical mechanics fail to predict this discreteness. Quantum theory, on the other hand, helps us to identify these permissible orbits for the electrons and calculate their energy levels. In the Bohr model for atoms, as taught in high schools, these are called the electronic energy levels in an atom or simply the atomic energy levels. According to Bohr's theory, an electron in the atom can absorb or emit light of a specific wavelength only, when it jumps from one permissible orbit into another. Let the electronic transition occur from a lower energy level $E_1$ to a higher energy level $E_2$, so that the energy gap $\Delta E = (E_1 - E_2)$. The absorbed wavelength $\lambda$ is obtained by using the formula $\Delta E = h\nu = hc/\lambda$, where the Planck's constant $h = 6.6 \times 10^{-34}$ $J.s$. This gives a dark line in the spectrum of the hydrogen atom. In this way, the observed series of lines in the hydrogen spectrum can be explained successfully.

Let us raise one question here (obviously attempted refutations are expected and welcome):

**Can we draw analogy of electrons (or photons) with neurons, the way we are observing as above?**

As you understand, our objective is to know about biological neurons for the ambition of producing a conscious machine. Incorporating consciousness into a machine is a complex task as rightly pointed out in ["Analysis and Implementation Strategy for Incorporating Consciousness into Machine Architecture" by C. N. Padhy and R. R. Panda, IEEE IACC-2009]. First, logical evaluation of consciousness is highly needed. Next, we have to formulate a set of instructions leading to this level of consciousness. Then only it would be proper to think of machine architecture. Of course, human body will be always the natural machine to imitate although there are plenty of deficiencies so far to be incorporated when we think of building our conscious artificial machine.

## 4. Developments in Physics up to the present state of the art leading hopefully to Consciousness

For many years until 1920 science had excluded 'consciousnesses' from the physical universe. However, a major paradigm shift in the 1920's from classical mechanics to quantum mechanics marked a break with that long tradition. Now the current interest in the foundation of 'consciousness' has led increasingly to the need to utilize quantum phenomena for the purpose of unfolding the physical reality. Hopefully, this will be able to change the complexion of the relationship between body and mind.

The principle of classical mechanics is that any physical system can be decomposed into a collection of simple independent local elements each of which interacts only with its immediate neighbors. Thus according to the ideas of classical physics, brain is a collection of massive system of parallel computers, one for each point in a fine grid of space time points that cover the brain over some period of time. In other words, the model of cellular automata might be helpful for the simulation work. Further, each individual computer would compute, record and transmit the values of the components of the electromagnetic and matter fields at the designated associated grid point. The major problem with beliefs and preconceived metaphor (called **"Sangskar"** in Hindu Philosophy) arises from the attempt to understand the connection of thoughts to brains within the framework of classical physics. Since the brain is a physical system, according to the precepts of modern physics the brain must in principle be treated as a quantum system. Later we shall see that even with the introduction of quantum concept, we arrive at some more contradiction, which signifies some more tools is required to understand 'mind' property of Nature. Nevertheless, for the time being let's point out one prime consideration of Quantum theory.

Quantum theory is unlike classical physics, in which a human consciousness is necessarily idealized as a no participatory observer, as an entity that can know the aggregate aspect of the brain. In quantum theory, the situation is subtler because there is a structural mismatch between a quantum mechanical description of a physical system and our perception of that system. This structural mismatch (we shall more elaborate on this topic at the end) is the basic entity of quantum theory and surely it opens up interesting possibility of representing mind/brain. This is because mind/brain should be thought as a combination of thought like and matter like aspect of a neutral reality. To describe the physical and chemical processes underlying brain action and existence of mind we thus need to move from monistic classical mechanics to a dualistic generalization. During the time when Heisenberg introduced the quantum mechanics, 'mind' was out of the purview of physics. But when this dualistic mechanics is applied to a human brain, it will hopefully account for the thought-like and matter-like aspects of the mind/brain system.

In the quantum description of Nature proposed by Heisenberg, reality has two different aspects:
1. One consists of a set of 'actual events' – these events form a sequence of 'happenings', each of which actualizes one of the possibilities offered by the quantum dynamics.
2. This consists of a set of 'objective tendencies' for these events to occur, these tendencies are represented as persisting structures in space and time.
   Our major research effort is to identify and correlate these persisting structures with the functionalities/events.

If we can correlate thoughts with high level quantum events in brains as suggested by von Neumann, Wigner and others, then we will be able to build-up a theory which will be dual aspect theory of the mind/brain, in the sense that it correlates the inner or mental aspects of mind/brain system with 'actual events' in Heisenberg picture of Nature. In this context, Bohr resolved the problem of reconciling the quantum and classical aspect of Nature by introducing the fact that, the only thing that is known to be classical is our description of our perceptions of physical objects. J. von Neumann and Wigner used this key insight into dynamical form by proposing that the quantum/classical divide be made not only on the basis of size, but rather on the basis of the qualitative difference in those aspects of Nature we call mind and matter (body).

Let us note the famous remark of Nobel Laureate Richard Feynman: "The theory of quantum electrodynamics describes Nature as absurd from the point of view of common sense and it agrees fully with experiment. So I hope you can accept Nature as she is –absurd".

The main ideas of Quantum Mechanics are as follows:

1. Energy is not continuous but in discrete units
2. The elementary particles behave both like particles and waves
3. The movement and behavior of these particles are inherently random.
4. It is physically impossible to know both the position and the momentum of a particle at the same time. The more precisely one is known, the less precise the measurement of the other is.
5. The microscopic world is different from our macroscopic world.

Brain and mind are two different entities as if one person (Brain) is always with his/her assistant (Mind). Think about the analogous situation when a person is about to do certain work, immediately this may be prohibited/supported by his assistant or mind with sufficient or physical interpretations are obviously available.

Einstein was one of the main players in the introduction of Quantum Mechanics. He himself explained the photoelectric effect with Planck's energy quanta, which were called 'photons' and introduced the concept of absorption and spontaneous and stimulated emission of radiation. In spite of all these tremendous achievements, Einstein always doubted the completeness of Quantum Mechanics-specifically the random nature of Quantum Mechanics and the spooky (alarming!) action-at-a-distance (non-locality) were the things he deeply rejected. His famous quote "Subtle is the Lord, but he does not play dice" demonstrates his attitude towards Quantum Mechanics. The deficiency of the Quantum Mechanics is more evident when we observe the following fact:

Let's raise the question "What are the ultimate physical limitations on computing power"? On using arguments based on Quantum Mechanics, H. J. Bremermann ("Optimization through Evolution and Recombination," in M. C. Yovits et al. (eds) Self organizing Systems, pp. 93-106, Spartan Books, Washington, D.C., 1962.) conjectured that no computer, either living or artificial can process more than $2 \times 10^{47}$ bits of information per gram of its mass per second. If it is true, computers like the size of the earth ($6 \times 10^{27}$ g) operating continuously for a period equal to the estimated age of the earth ($10^{10}$ years) could then process fewer than $10^{93}$ bits.

From the above discussion one thing we may realize that mass, information processing capacity and energy are all synonymous and hence interchangeable.

The points, which run against the above conjecture and hence incompleteness of Quantum Physics could be the following:

1. It is much less than the number of possible sequences of moves in simple chess game which has been estimated at $10^{120}$.
2. Further, human brain supposedly the most complex machine in the universe has 100 billion ($10^{11}$) neurons. Scientists already know that neurons convey information via electrical spikes, and each neuron interconnects with hundreds of other neurons via, on average 10 thousand connections or synapses. Thus each brain has about thousand trillion ($10^{15}$) synapses. Combining all the brains of living organisms to date must be much bigger than the so far processed bits of $10^{93}$ as per Quantum Mechanics.

Now let us elaborate on the precision of wiring between neurons in the formation of networks and modules. It is to be noted that neurons are not simply interconnected with any or all other neurons; rather they make precise connections with a subset of cells and form networks that process information. Obviously the pathways that convey information and networks that process them are at the core of brain's function. Next we elaborate on Cognition and mind, which are essentially believed to be the function of some network in the brain:

One thing must be clear that one or several networks must be able to perform the following functions from the point of view of cognitive sense:

1. Parasitic behavior, meaning the tendency to acquire resources for self-survival without concerning other's existence. In other words, the aggregate behavior makes a being autonomous
2. Symbiotic Behavior, meaning the tendency to associate peers for the purpose of strengthening and smoothening the survival process
3. Self-referral behavior meaning the tendency to recursively referencing the self (the same network or/and a collection of networks). In other words, performing the actions to identification and protection for the self.
4. Reproductive behavior meaning that it plays a role for multiplication, a role in populating self-species type for the purpose of competing with other species in the struggle for survival.

The thing that is of prime importance is that all the above cognitive behaviors we observe must be true from the microscopic world of cell level to the macroscopic world of organ or functional or behavioral level. Obviously our nervous system plays a critical network role for the above kind of integration from the micro to macro level.

In a nutshell let us revisit the various significant points as the following:

**Newtonian Mechanics:**

What are the main features of Newtonian Mechanics?

1. It is just Euclidian Geometry (3-D Space) Here time has not been taken in to consideration. As if both Event and observation are taking place simultaneously in space and time. Thus absolute time and space have been considered.
2. The three fundamental dimensions of Mass (M), Length and Time are considered absolute.
3. Newtonian mechanics is completely deterministic.
4. Newtonian Mechanics applies to macro system.
5. Event and observation are equivalent with respect to energy and momentum.

**Einstein's Relativistic Mechanics (Special)**

1. All the three fundamental dimensions (M, L, T) are relative except the rest mass $m_0$, rest length $l_0$ or rest time $t_0$. It is the so-called Minkowski's geometry.
2. It applies to the inertial frame of reference. Any particle is at rest or in uniform or uni-directional motion.
3. Equivalence of event and observation with respect to the observer.
4. It is also deterministic.
5. Therefore it can only be applicable to macro system where the changes are negligible.

**Quantum Mechanics**

1. QM incorporates the non-equivalence of event and observation in the light of Heisenberg's uncertainty principle
2. It applies to the non-inertial frame of reference. That is, velocity and direction continuously changing with respect to time.
3. It is non-deterministic.
4. It can be applicable to micro system.
5. It applies to non-linear geometry.

What is lacking in all the above formalisms to describe the mechanics of a physical system concerning quantum particles? By quantum particle we mean any particle whose mass is $\approx$ less than $10^{-5}$ gram (Plankian mass).

The last but not the least significant point, which is lacking is to incorporate absolute universal units of mass, length and time which must be incorporated to a theory to give it to a complete and deterministic form. This is quite logical from the viewpoint of precision of any measurement theory.

## 5. A Theory of Quantum Consciousness

Here we build up a qualitative as well as quantitative concepts of the widely used term "consciousness" from the most fundamental level of quantum reality, as is manifested in nature.

The mathematically formulated quantum consciousness parameter (QCP) has been used to explain analytically the basic perceptual and conceptual details of the interactions of the observer and the observable via the process of observations done by communicating signals, which after interacting with the observable bring (back) information about the same to the observer.

The entire treatment has been constructed essentially on the basis of a new deterministic interpretation of conventional quantum mechanics which in our view actually a complimentary picture of the existing formalism and in no sense should be miss interpreted as a contradiction of the existing one. In other words, the deterministic status, which has been established earlier in the paper (physical Essays volume 8, number 4, December 1995), provides a fuller description of the reality of the quantum world in terms of a broader duality. We have to consider the duality of the probabilistically perceptible dynamical variables and the hidden form of deterministically computable variables. Thus keeping in mind the necessity of mathematically formulated definition of consciousness, we have constructed the theory in a very generalized form starting from fundamental quantum entities. This will obviously fuel the future rapid advancement of science and technology because the more complex macro bodies are nothing but ensembles of quantum particles.

Before going into the theory part, some important revelations are the following:

1. All the individual entities in the universe are constantly in the act of interaction with each other in multifaceted manner strictly obeying the laws of physics both in the micro and macro levels, no matter whatever be the space time separation between them (Ernst Mach's Principle). Therefore it can be concluded that consciousness is the most fundamental universal parameter, which is inseparably associated with all material existence, no matter whether living or non-living.

2. Outline of treatment in this section is an extension of the idea of quantum consciousness (associated with quantum entities) from the simplest to the more complex system of quantum (ensembles). Then the elementary idea has been applied, in an extended form to have a more generalized and explicit view of "System consciousness", both in the areas of biological science and man-made machines. Because some forms of a "self-organized automata" is a common feature, manifested in both living and nonliving systems, which perform certain self-operational functions, which are actually the result (product) of the quantum interactions constantly going on in their quantum level of composition. The simplest examples clarifying the above notion are the atom (non-living system) and the cell (living system). Both of them having their own distinct built in system consciousness, which are the guiding and governing factors throughout their internal and environmental (external) responses.

3. The quantum consciousness parameter (QCP) and its relation with the four other dynamical variables, viz. relativistic energy, momentum, time period and wave length have been established in the following theory part clearly describing the underlying physical significances and distinctively different properties of QCP from other parameters (energy, momentum, etc).

However, complimentary picture of the new idea with the existing one is also discussed in details. Thus it is foreseen that QCP would provide a mathematical tool for the scientists to apply the same effectively in all the areas of quantum mechanics especially in the deeper understanding of quantum mechanics, quantum information theory and quantum computation. Other significant breakthrough could be achieved in technology by constructing mathematical working models, based on the concepts of QCP and applying them to the challenging problems in the areas of molecular biology, medical and medical instrumentation, genetics, neuro-science, biotechnology, bio-informatics and psychology. The most important aspects of the future QCP-based nano-technology would be its high level precision factor due to its inbuilt quantum deterministic status as shown in the theory part. A multi disciplinary scientific research programme is solicited to incorporate the QCP concept effectively in modern nano-technology in almost all the areas of modern science. We have also undertaken the task of providing the basic algorithms as a preliminary working tool. Essentially, this basic principle could be used in the different areas of applications mentioned above according to their specific requirements, for different data based computation. Before going into the theory part we cannot resist the temptation of quoting from the writing of two great physicists, Prof. Max Born (Nobel laureate) and Prof. Franck Wilckzek (Nobel laureate 2004) both of them Nobel Prize winner in physics. Their views about the future of quantum physics clearly indicate that the task of QM emerging as a unitary science, embracing all the areas of natural philosophy has been achieved. The problems of consciousness will remain/ yet to be solved before such a goal can be reached. Nothing can be excluded from our interest of research in our quest for deeper knowledge. The enigma of both inert matter and those of the living material entities can possibly be explained in terms of the more generalized and omnipresent existence of quantum consciousness as a unitary science.

**Theory:** We have mentioned above that a hidden deterministic theory (vide Physics Essay Vol-8 & No. 4) of QM is the foundation of the theory of quantum consciousness (QC).

Heisenberg's Uncertainty principle is introspected and reevaluated in the light of the non-infinite and non-zero quantized energy and momentum limits lending to the following results:

i) Functional qualitative relations are established between E and $f(\Delta E)$ (where f stands for function) and p and $f(\Delta p)$ where E and p are respectively the actual values of the energy and momentum of a measured particles. $(\Delta E)$ and $(\Delta p)$ stands for their respective uncertainties due to measurement: $E = f(\Delta E)$ and $p = f(\Delta p)$. The equation for E is given in [Physics essays, vol.8, no.4] as for the readers as a reference, which would be required henceforth time to time.

ii) This reveals the hidden intrinsic deterministic characteristics of QM and dynamical variables (E, p, $\lambda$, $\tau$), which are mathematically understandable and computable. It is only their probabilities that can be measured by observation.

It is, therefore ethically required to replace the wave characteristics terms $\lambda$, $\tau$ by other terms befitting the new deterministic formalism. A comparative picture between the two interpretations are given below:

i) Conventional view: E=h.ν where ν is the frequency and h is Planck's constant. ν = $1/\tau_c$ where $\tau_c$ is the Compton time period and $\tau_c = h/m_v C^2$.

   Where, E = $m_v C^2$ is the relativistic energy of a particle and $m_v = \dfrac{m_0}{\sqrt{1-\dfrac{v^2}{c^2}}}$ where $m_0$ is the rest mass and $m_v$ is the relativistic mass of the particle traveling with velocity v relative to the observer.

ii) De Broglie's wave equation is p=h/λ where p=$m_v v$ where $m_v = \dfrac{m_0}{\sqrt{1-\dfrac{v^2}{c^2}}}$ is the relativistic momentum of a particle and λ is De Broglie's wavelength. The De Broglie time period $\tau_\delta$ =$h/m_v v^2$. This probabilistic interpretation is based on Max Born's idea of wave of probability and Schroindinger's wave function ψ, which is physically signified as the amplitude of the probability.

The new concepts of hidden determinism are based on the idea of sharp discreteness:

Here the term λ is replaced by L=h/$m_v v$ and v is simply the particle velocity (not group velocity $v_g$) and L is spatial dimension in the direction of its trajectory of motion. This term $\tau_\delta$ is replaced by T=h/ $m_v C^2$. This is the time required by the particle to traverse its own length. Therefore the non-observable velocity of propagation is expressed V=L/T=$C^2$/v ≥C (where v is the group velocity in wave interpretation whereas here in our new interpretation v is the particle's relativistic velocity) which was termed as a phase velocity in the conventional wave interpretation. According to S.T.R surely the velocities greater than C will fetch only imaginary mathematical existence because the term $1-\dfrac{v^2}{c^2}$ becoming negative.

Conventional wave mechanics in quantum theory only attributes a mathematical significance to V as the phase velocity of a group of unidirectional waves. But in the light of the new deterministic scenario, it is suggested that V can be defined as the velocity of propagation of the quantum consciousness of quantum particles. In the present context a quite similar situation can be cited where P.A.M Dirac faced the problem of providing a reasonable physical explanation for the negative roots of his formulated quantum-relativistic energy equation E=$\sqrt{p^2 C^2 + m_0^2 C^4}$ for, at that time, nothing like "Negative Energy States" was known to the physicists. The reality of Dirac's prediction of antiparticles was later experimentally confirmed.

Therefore, it should not be taken for granted that a physically non-perceptible entity, which is only mentally conceivable, does not exist as a physical reality. In the case of quantum consciousness the reason behind its non-manifestation is that its velocity of propagation V is greater than C. Knowing that the mind and body is an intimately integrated system as a whole it is reasonable to incorporate quantum consciousness as the most fundamental physical parameter, which exists as a reality in all material entities whatsoever. It is only the levels of consciousness depending upon the degrees of self-organized automata and complexity of structure in isolated systems, which makes the quantitative as well as qualitative differences in the manifestation of their multifaceted awareness, due to their interactions with other systems. Thus the present paper would throw light on a unified notion or concepts of matter, energy, mind and consciousness and will give a directive in the construction of a future unitary science, which would enable us to have a deeper as well as broader insight of our phenomenal nature. As consciousness means knowledge, it is quite reasonable to enunciate quantum consciousness as the knowledge about the states of a quantum particle, relative to an observer. Here it is to be understood and kept in mind that quantum consciousness parameter QCP in its dimensional structure must involve all the four dynamical variables (energy, momentum, time and space) in their uncertainty free, pure form to attribute totality as well as purity of the quantum information contained in it (Please recall here about the **'Sat-Chit-Ananda'** as explained in the introduction).

The quantitative equivalence of $K_\varphi$(QCP) of a quantum particle is conceptualized here as its impulse of relativistic energy relative to an observer. The term impulse is dimensionally the same as momentum and the preposition "of" implies product, (so for a non zero rest mass particle) we can write,

$$K_\varphi = (mv)(mc^2) = m^2vc^2 \text{ Or } \{K_\varphi\}_{m \neq 0} = m^2vc^2$$

Similarly for zero rest mass particles e.g photons

$$\{K_\varphi\}_{m \neq 0} = \left(\frac{hv}{c}\right) \times hv = \frac{h^2v^2}{c} = m^2c^2$$

Thus it is significant that our theory would serve as the philosophical foundation of the more versatile unitary theory of QC. Unitary in the sense, that it unifies the non-observable aspect of QM with its observable counterpart, which was lacking in QM as a relevant complimentary.

Also as $\{K_\varphi\}_{m \neq 0} = m^2c^2 = h^2/LT$, the space and time are therefore parameters automatically included in $K_\varphi$.

We can write the energy equation of a particle in terms of its uncertainty in measurement as

$$E = \left\{E_p^{\frac{1}{2}}(4\pi E_i)^{\frac{1}{2}}\right\} \times \left(\sqrt{\Delta E} - E_0\right) \dots\dots\dots\dots\dots\dots(4)$$

Where $E_p$ is Planckian energy= $2\pi\sqrt{(hc^5/G)}$, h denotes Planck constant, c is velocity of light and G is Newtonian gravitational constant. $E_i$=$hH_0/2$ $H_0$ in Hubble's cosmological parameter, considered here as a constant is the min non-zero quantized energy derived from the scale of the universe. $E_0$ is the geometric mean of $E_p$ and $E_i$.

Therefore $E_0 = (E_p * E_i)^{1/2}$

As theoretically the non zero value of $\Delta E$ cannot be lesser than $E_i$ i.e. $\Delta E = E_i$ we can substitute $\Delta E = nE_i$ where n is natural number. In the equation 4 and write it in the form given below…

$$E = \left\{E_p^{1/2} (nE)_i^{1/2} + (4\pi E_i)^{1/2} * (nE)_i^{1/2}\right\} \dots\dots\dots\dots\dots\dots (4a)$$

Or $E = \left\{n^{1/2} E_0 + (4\pi)^{1/2} * (nE)_i^{1/2}\right\} - E_0$

From equation (4) $E_i = E_0 * \left(\frac{E_0}{E_p}\right) = \mu E_0$, $\mu = \frac{E_0}{E_p}$ which is non-dimensional constant.

Therefore (4a) can be expressed as

$$E = \{n^{1/2} E_0 + (4\pi)^{1/2} \mu * (n^{1/2} E_0)\} - E_0$$

$$E = \{1 + (4\pi)^{1/2} \mu\} n^{1/2} E_0 - E_0$$

Writing the n.d constant $\{1 + (4\pi)^{1/2}\mu\} = \beta$ the simplest form of (4a) becomes

$$E = \beta \{n^{1/2} - 1\} E_0 \ldots\ldots\ldots\ldots\ldots\ldots(4b)$$

Returning to equation of QCP $K\varphi = m^2 v c^2 = E^2(v/c^2)$

Substituting (4b) in 1 we get $K\varphi = \beta \{n^{1/2} - 1\} E_0^2 (v/c^2)$

Considering the value of the constant β we get $\beta = \{1 + (4\pi)^{1/2}\}$ Where $\mu = 10^{-31}$ which is very small quantity, so ignoring the second term in β we can take its value as 1.

Thus $K\varphi = \{n^{1/2} - 1\} E_0^2 (v/c^2)\ldots\ldots\ldots\ldots(4c)$

The direction of Kφ is the same as that of v.

Studying (4c) we can see that the magnitude of Kφ is extremely small excepting for every large values of n because $E_0 \approx 10^{-16}$ erg and even for velocity very close to $c$.

$v/c^2 \approx 10^{-10}$ cm$^{-1}$ sec for n>>1 one can also write $K\varphi = n E_0^2 (v/c^2)\ldots\ldots\ldots\ldots(4d)$

Similarly, for the zero rest mass particles e.g. photons,

Thus it is significant that QCP = $m^2 v c^2$ as the deterministic hidden variables theory is the foundation of consciousness theory.

But, in the light of the new deterministic picture, it is suggested that V can be defined as the velocity of propagation of the quantum consciousness of quantum particle traversing its own length. Therefore the non-observable velocity of propagation is expressed as $V = \dfrac{k_\varphi}{m^2 c^2}$

Which was termed as phase velocity in the conventional wave (inter pertain). According to S.T.R velocities greater than $c$ can have only imaginary mathematical existence because the term becomes negative. Conventional wave mechanizes in quantum theory only attribute (a) mathematical significance to as the phase velocity, of a group of unidirectional waves. This rational directive in formulating QCP a depending upon the degrees of self-organized automata and complexity of structure isolated systems, which makes the quantitative as well as qualitative differences in the manifestations of their multifaceted awareness due to their interactions with other systems.

In the case of quantum consciousness the reason behind its non-manifestation is that its velocity of propagation v is greater than $c$.

Therefore, it should not be taken for granted that physical entity which is mentally conceivable non-perceptible does not as a physical reality.

Knowing that the mind and body is an intimately integrated as a soul, it is reasonable to incorporate quantum consciousness as the most fundamental physical parameter, which exist as a reality in all material entities irrespectively. It is only the levels of consciousness which matters for different species; the present paper would throw light on a unified notion or concepts of matter, energy, mind, consciousness which would enable us to have deeper as well as broader view of our phenomenal nature.

As consciousness means knowledge, it is quite reasonable to enunciate quantum consciousness as the knowledge about the state of a quantum particle, relative to an observer. Here it is to be understood and kept in mind that the quantum (energy, momentum, time and space) is in their uncertainty free, pure form, to attribute totality as well as purity of the quantum information contained in it.

## 6. Conclusion and Future Research Directions

In this paper consciousness has been dealt from various viewpoints. First it has been conceptualized as the momentum times energy and as a result the force of consciousness has been derived. On introducing fractal geometry, it has been conceived that more prosperous species (with regards to consciousness) might be having higher fractal dimension in the phase space of electroencephalographs representing the higher consciousness. Then systematic gradual advancements of Physics have been enumerated leading to the complimentary theory of quantum consciousness.

Thus it is foreseen that QCP would provide a mathematical tool for the scientists to apply the same effectively in all the areas of quantum mechanics especially in the deeper understanding of quantum mechanics, quantum information theory and **quantum computation.** Other significant breakthrough could be achieved in technology by constructing mathematical working models, based on the concepts of QCP and applying them to the challenging problems in the areas of molecular biology, medical instrumentation, genetics, neuro-science, biotechnology, bio-informatics and psychology. The most important aspects of the future QCP-based nano-technology would be its high level precision factor due to its inbuilt quantum deterministic status as shown in the theory part. A multi disciplinary scientific research programme is solicited to incorporate the QCP concept effectively in modern nano-technology in almost all the areas of modern science. We have also undertaken the task of providing the basic algorithms as a preliminary working tool. Essentially, this basic principle could be used in the different areas of applications mentioned above according to their specific requirements, for different data based computation.